\documentclass[aps,prl,preprint,groupedaddress,showpacs]{revtex4}
\usepackage[dvips]{graphicx}
\usepackage{subfigure}
\begin{document}

\title{Resonance beating of light stored using atomic spinor polaritons}

\author{Leon Karpa}
\author{Frank Vewinger}
\author{Martin Weitz}
\affiliation{Institut f\"ur Angewandte Physik der Universit\"at Bonn, Wegelerstr. 8, D-53115 Bonn, Germany}

\date{\today}

\begin{abstract}
We investigate the storage of light in atomic rubidium vapor using a multilevel-tripod scheme.
In the system, two collective dark polariton modes exist, forming an effective spinor quasiparticle.
Storage of light is performed by dynamically reducing the optical group velocity to zero.
After releasing the stored pulse, a beating of the two reaccelerated optical modes is monitored.
The observed beating signal oscillates at an atomic transition frequency, opening the way to
novel quantum limited measurements of atomic resonance frequencies and quantum switches.
\end{abstract}


\maketitle

\section{}
Electromagnetically induced transparency (EIT) is a quantum
interference effect that allows for the transmission of light
through an otherwise opaque atomic medium
\cite{10.1103/RevModPhys.77.633}. More specifically, an optical
``control'' beam is used to dress the medium to allow for the
transmission of optical pulses from an optical ``signal'' beam.
Media exhibiting EIT have remarkable properties, as very low group
velocities
\cite{Nature.397.594-598(1999),PhysRevLett.82.5229,PhysRevLett.83.1767}.
Associated with a slow light propagation are quasiparticles, the
so-called dark polaritons, which propagate through the medium with
the speed of the group velocity. The group velocity can even be
reduced to zero in a controlled and reversible way, which allows
for the storage of light. Notably, also non-classical states of
light, as single-photon or squeezed states, have been stored and
retrieved in atomic ensembles in a series of impressive
experiments
\cite{Nature.409.490-493(2001),PhysRevLett.86.783,10.1038/nature04327}.
Applications in quantum information science have been suggested
\cite{PhysRevLett.88.243602}. However, a quantum bit
$\left|\psi\right\rangle=\alpha\left|0\right\rangle +
\beta\left|1\right\rangle$ has two basis states, so that qubits
cannot be stored in a single atomic interaction region when using
conventional EIT media consisting of atoms with a $\Lambda$-type
coupling scheme, which exhibit a single dark state. In a recent
remarkable experiment, photonic quantum bits have been stored in a
superposition of two spatially separated EIT interaction regions
\cite{10.1038/nature06670}. It is however clear that such a
spatial separation makes the system sensitive to (inhomogeneous)
magnetic stray fields and optical path length fluctuations. Thus
multi-component dark polaritons are of interest, as these should
allow for the storage of photonic qubits in a single atomic
ensemble.

Our work builds upon an ensemble of atoms in a tripod-type level
configuration, where an internal state subspace immune to
spontaneous decay spanned up by two orthogonal dark eigenstates
exists \cite{PhysRevA.54.1556,PhysRevLett.91.213001}. It has been
suggested that onto this two-dimensional decoherence-free subspace
qubits can be encoded, and the system should allow for interesting
quantum manipulation \cite{PhysRevA.65.032318} and geometric
phases \cite{Phys.Rev.A77.022306}. Other theoretical work has
shown that in an ensemble of atoms with tripod level structure
('tripod medium') two optical modes with slow group velocity can
exist \cite{PhysRevA.70.023822}.

Here we demonstrate the reversible storage of amplitude and
relative phase information of two optical fields in rubidium
atomic vapor. Effectively, slow light spinor states are stored and
coherently retrieved from the atomic ensemble. It is suggested
that the coherent propagation of light through the tripod medium
can be described by a dark polariton with an internal two-level
structure, whose group velocity can be dynamically varied to zero,
thereby mapping photonic qubits onto an appropriate spin coherence
information. We furthermore show that after the storage procedure,
the observed beat frequency of the two reaccelerated optical
fields is determined by the energetic splitting between two ground
state sublevels. The difference frequency of the emitted beams
remains synchronized to an atomic resonance frequency, so that the
operation of our experiment can be characterized as a
stored light atomic clock.

Before proceeding, we point out that the two-dimensional
dark subspace of the tripod medium has an equal number of degrees
of freedom as a free photon of given frequency in the quantum
state $\left|\psi\right\rangle=\alpha\left|H\right\rangle +
\beta\left|V\right\rangle$, where $\left|H\right\rangle$ and
$\left|V\right\rangle$ denote two polarization eigenstates.
Correspondingly, tripod media can be used to slow down (and store)
photons of arbitrary polarization.

\begin{figure}[h!]
  \begin{center}
    \includegraphics [width=3cm]{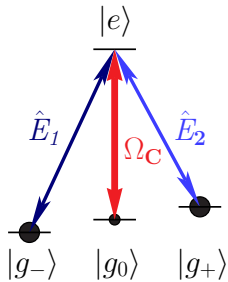}
  \caption{\label{fig:tripod} Simplified scheme of relevant levels, as used for the theoretical
  calculations described in the text. The levels $\left|g_{0}\right\rangle$
  and $\left|e\right\rangle$ are coupled by one strong control field with the Rabi frequency
  $\Omega_{C}$ while the two signal fields described by the quantum field operators $\hat E_{1}$
  and $\hat E_{2}$ drive transitions between $\left|g_{-}\right\rangle$ and $\left|e\right\rangle$
  and $\left|g_{+}\right\rangle$ and $\left|e\right\rangle$ respectively. The dots are to indicate
  the ground state populations.}
  \end{center}
 \end{figure}
For a simple model of dark state polaritons in a tripod medium, consider the level scheme
of Fig. \ref{fig:tripod} with three stable ground states
$\left|g_{-}\right\rangle$, $\left|g_{0}\right\rangle$ and $\left|g_{+}\right\rangle$
and one spontaneously decaying excited state $\left|e\right\rangle$.
An ensemble of such four-level atoms is resonantly coupled to a strong,
classical "control" laser field connecting states $\left|g_{0}\right\rangle$ and
$\left|e\right\rangle$ with Rabi frequency $\Omega_{C}$ and two weak, quantum signal
fields described by field operators $\hat E_{1}(z,t)$ and $\hat E_{2}(z,t)$ connecting
the quantum states $\left|g_{-}\right\rangle$, $\left|e\right\rangle$ and
$\left|g_{+}\right\rangle$, $\left|e\right\rangle$ respectively, where we
assume equal atom-field coupling constants $g$ for both transitions.

In the absence of a signal field, the combined atom-light system
is optically pumped into a state described shortly by
$\hat{\rho}_{vac}=$$\frac{1}{2} ( \left|g_{-}\ldots
g_{-}\right\rangle\left\langle g_{-}\ldots g_{-}\right| + \left|
g_{+}\ldots g_{+}\right\rangle\left\langle g_{+}\ldots
g_{+}\right|)$$\left|0\right\rangle_{1}\left|0\right\rangle_2\left\langle
0\right|_2 \left\langle 0\right|_1$ determined by an incoherent
superposition of half the atoms in state
$\left|g_{-}\right\rangle$ and half in state
$\left|g_{+}\right\rangle$, while both signal fields are in the
vacuum state. The state described by this density matrix is named
the polariton vacuum state (in analogy to the work reported in \cite{PhysRevLett.84.5094}). Petrosyan and Malakyan showed that in
such a medium with N uniformly distributed atoms per unit length
two dark signal fields propagate, both with group velocity
$v_{g}=c/\left( 1+\frac{N\cdot g^{2}}{2\Omega_{C}^{2}} \right)$
\cite{PhysRevA.70.023822}. Starting from the solutions found by
Fleischhauer and Lukin for a three-level medium
\cite{PhysRevLett.84.5094}, we can due to the linearity of the
Hamiltonian for small signal field strengths construct the
corresponding dark polariton states for the tripod medium:
\begin{equation}
    \hat\Psi_{\pm}(z,t)=\cos\Theta\cdot\hat{E}_{1,2}(z,t)-\sin\Theta\sqrt{N/2}\;\cdot\hat{\sigma}_{\pm,0}(z,t)
\end{equation}
where $\tan\Theta=\sqrt{N/2}\cdot g/\Omega_{C}$ and
$\hat{\sigma}_{\pm,0}=\frac{2}{N}\sum_{j=1}^N
\left|g_{\mp}\right\rangle \left\langle g_{0} \right|$ are
spin-flip operators. Further, we can define
\begin{equation}
    \hat{\Psi}(z,t)=\alpha\hat{\Psi}_{+}(z,t)+\beta\hat{\Psi}_{-}(z,t),
\end{equation}
as the two-state polariton operator. The group velocity $v_{g} = c \cos^2\Theta$  can dynamically be varied, so that we expect that such "spinor polariton states" can be coherently manipulated analogous to usual slow and stored light experiments.

A one-polariton state can be generated using this operator, which yields $\hat\Psi^{\dagger}\hat\rho_{vac}\hat\Psi$. It is interesting that the operators $\hat\Psi_{\pm}$ (as well as the qubit state $\hat\Psi=\alpha\hat\Psi_{+} + \beta\hat\Psi_{-}$) describe pure polariton states, though they act on an incoherent ensemble. Using these polariton operators repeatedly, we can also construct different polariton Fock states \cite{PhysRevLett.84.5094} or coherent polariton states \cite{Las.Phys.15.3(2005)}, which are expected to be closest to the actual states produced in our experiment that is carried out with classical optical fields.

\begin{figure}
     \centering
       \subfigure[]{
           \label{fig:cell}
           \includegraphics[width=0.49\textwidth]{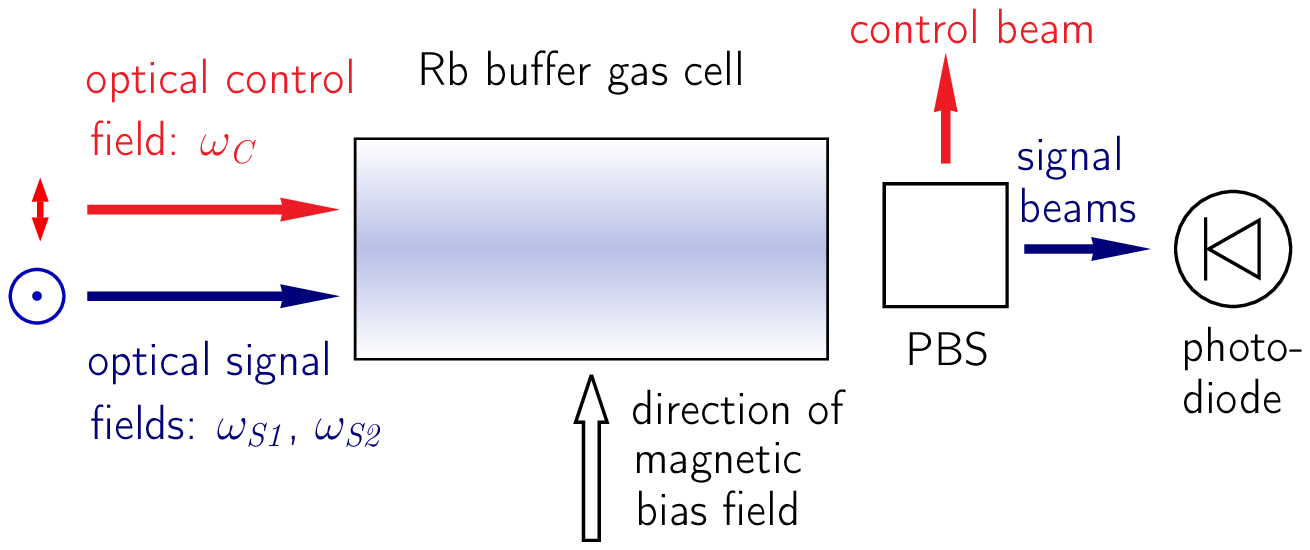}}
        \centering
     \subfigure[]{
           \label{fig:LambdaM}
           \includegraphics[width=0.45\textwidth]{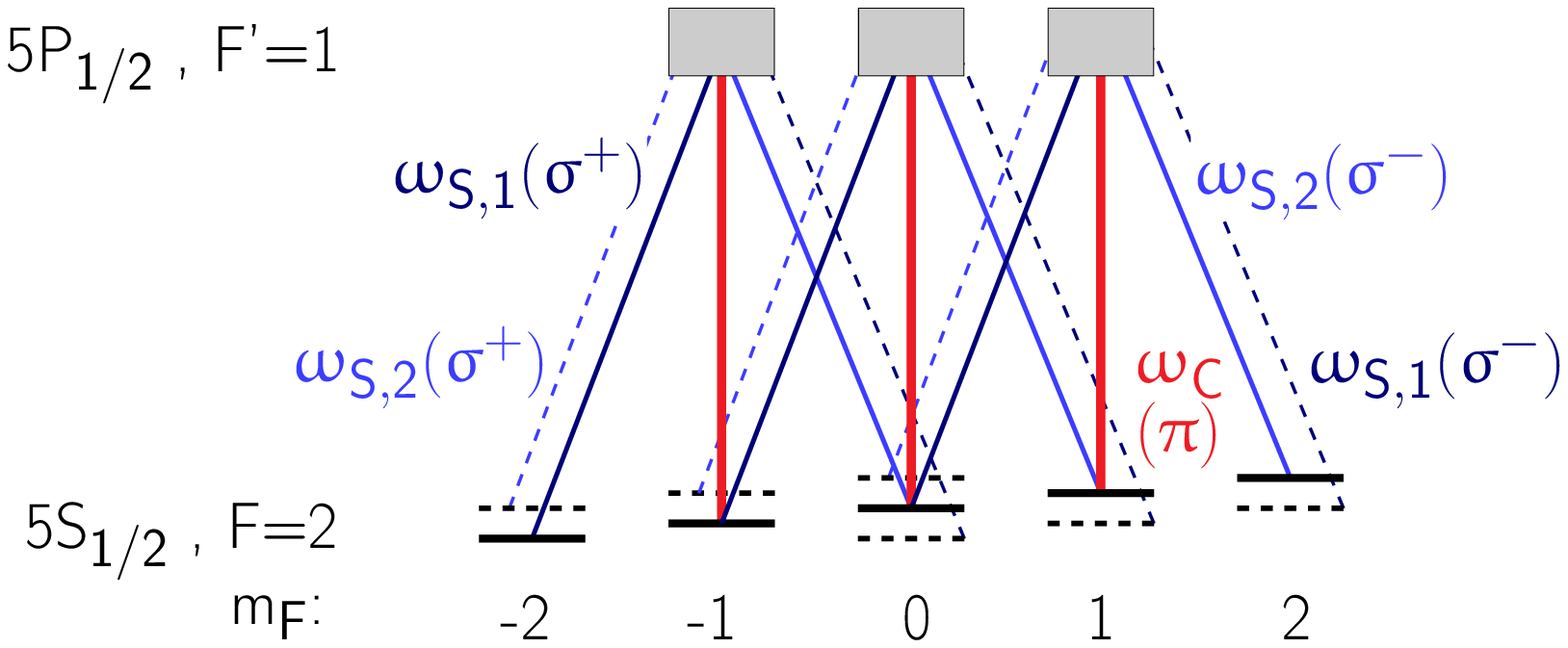}}
     \caption{\textbf{(a)} Scheme of the experimental setup. The input quantum state, encoded in a superposition of the two signal fields with optical frequencies $\omega_{S1}$ and $\omega_{S2}$, is reversibly stored in a rubidium buffer gas sample. The optical control field with frequency $\omega_{C}$ has an orthogonal linear polarization with respect to the signal fields. The signal fields and the control field are spatially overlapped upon entry into the rubidium cell.
      \textbf{(b)} Full scheme of the rubidium levels addressed by the two signal fields and one control field (solid lines). Also shown are the off-resonant circularly polarized components of the signal beams, i.e. the component with frequency $\omega_{S2}$ and $\sigma_+$-polarization and that with frequency $\omega_{S1}$ and $\sigma_-$-polarization (thin dashed lines).}
     \label{fig:setup}
\end{figure}

Our experimental setup is similar as described previously
\cite{New.J.Phys.10.045015(2008),2006NatPh.2.332K}, and a
schematic is shown in Fig. \ref{fig:cell}. A 50 mm long rubidium
vapor cell filled with 10 torr of neon buffer gas placed into a
magnetically shielded region is heated to approximately
$100\,^{\circ}\mathrm{C}$. Both optical signal fields and the
control field are derived from the same grating stabilized diode
laser, locked to the $F=2 \longrightarrow F'=1$ component of the
rubidium D1-line near 795 nm. Its emitted radiation is split into
two beams, both of which pass an acousto-optic modulator. One of
the paths is used to generate the optical signal fields and the
corresponding modulator is driven with two distinct radio
frequencies to generate the two signal beam frequencies
$\omega_{S1}$ and $\omega_{S2}$. The other modulator is driven
with a single radio frequency to produce the optical frequency of
the control field. The beam paths are overlapped, sent through an
optical fiber, and before entering the rubidium apparatus expanded
to a 2 mm beam diameter.

A full level scheme of our rubidium system is shown in Fig.
\ref{fig:LambdaM}. The $F=2$ ground state has 5 Zeeman components,
while the $F'=1$ excited state has 3 sublevels. As in the
simplified scheme of Fig. \ref{fig:tripod} the conclusion remains
valid that two dark states exist, as the characteristics of a
tripod medium. The rubidium cell is subject to a 150 mG
homogeneous magnetic bias field directed transversely to the
propagation axis of the optical beams. The control beam has
$\pi$-polarization, so that in the absence of signal beam light
the atoms will be pumped into an incoherent superposition of the
two Zeeman ground states $m_{F} = -2$ and $m_{F} = 2$
respectively. Both signal fields have the same
$\sigma_{+}\sigma_{-}$ (i.e. linear) polarization. Signal field 1
(2) is tuned to a frequency
$\omega_{S1}=\omega_{C}+\frac{\mu_{B}\cdot B}{2}$
($\omega_{S2}=\omega_{C}-\frac{\mu_{B}\cdot B}{2}$), where
$\omega_C$ denotes the optical frequency of the control field.
Thus, the $\sigma_+$-component of signal field 1 along with the
$\sigma_-$-component of signal field 2 and the $\pi$-polarized
control field give the usual coupling scheme of a tripod system.
Note that the $\sigma_{-}$-polarized component of field 1, and the
$\sigma_{+}$-polarized component of field 2 (indicated as dashed
lines in Fig. \ref{fig:LambdaM}) provide decay channels, so that
we expect the dark states to show a non-zero leakage; i.e. strictly speaking we have gray states \cite{loss_comm}. After passing
the rubidium cell, control and signal beams are separated with a
polarizing beam splitter and the total intensity of both signal
fields is detected on a photodiode.

\begin{figure}
     \centering
      \subfigure[]{
           \label{fig:darkres}
           \includegraphics[width=0.3\textwidth]{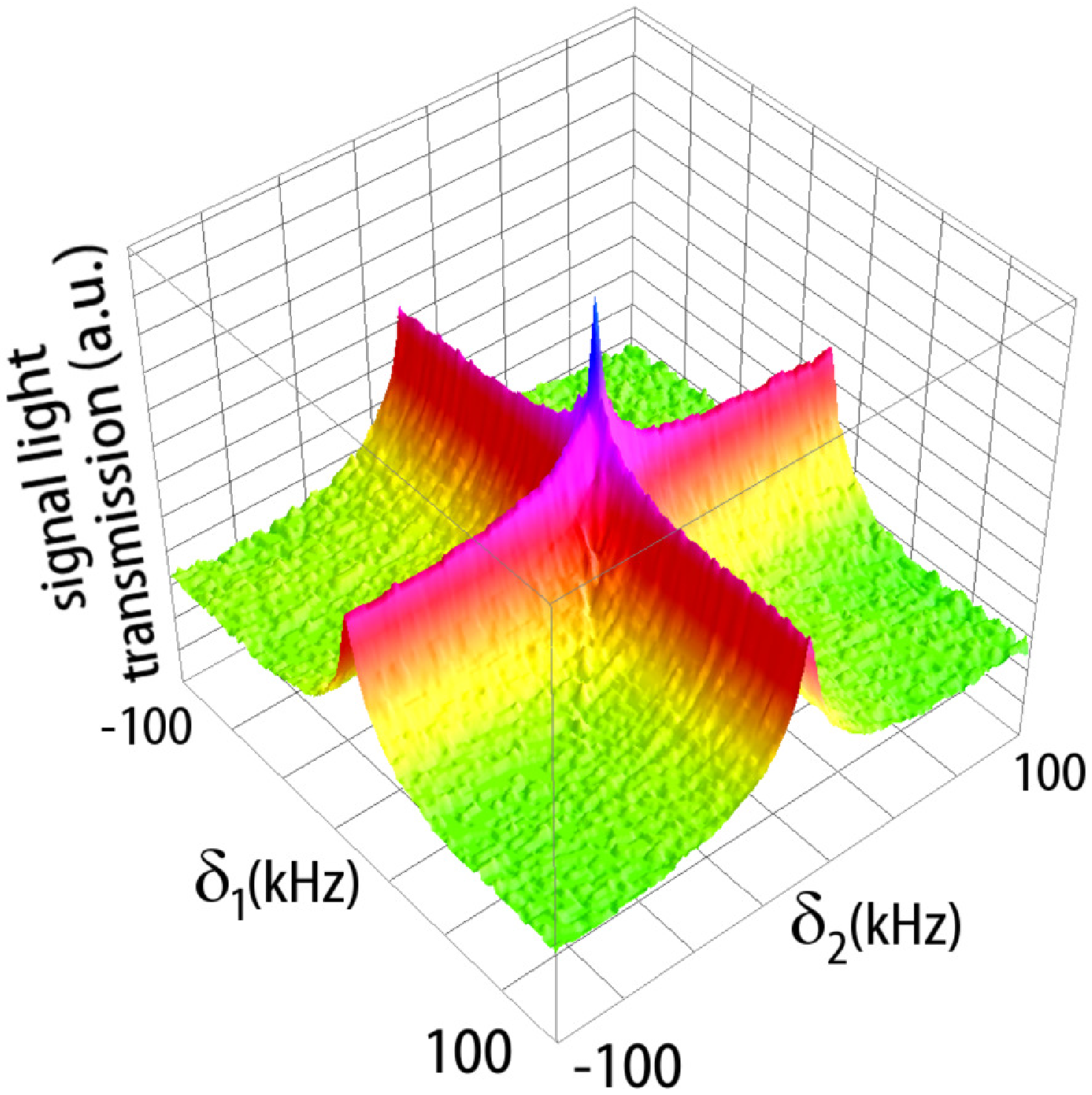}}
\hspace{1cm}
        \centering
     \subfigure[]{
           \label{fig:storedlight}
           \includegraphics[width=0.3\textwidth]{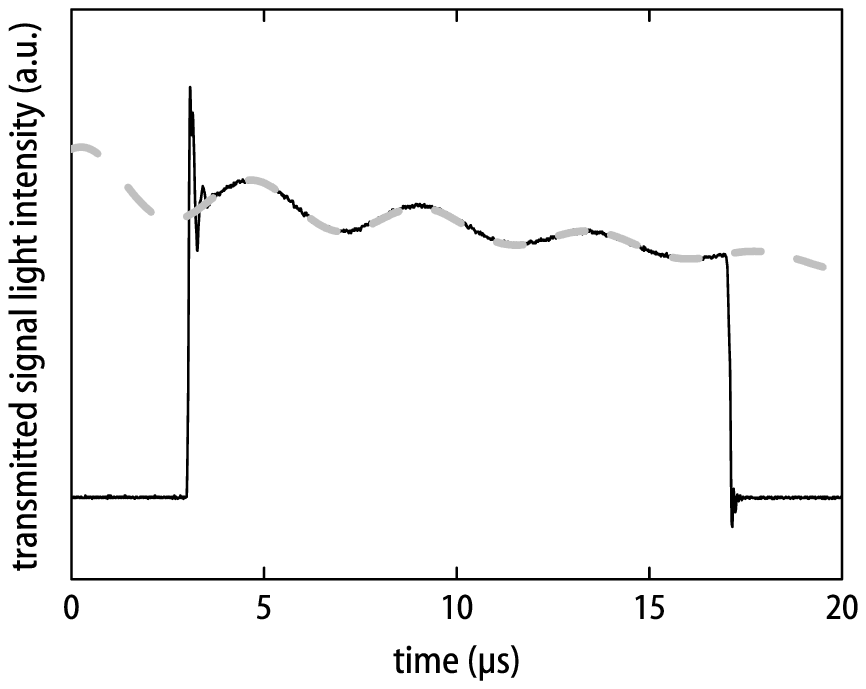}}
     \caption{\textbf{(a)} Total signal beams transmission signal (dc photodiode signal)
     averaged over 10 scans
     versus the two-photon detunings $\delta_{1}$ and
     $\delta_{2}$.  A visible small modulation near the line determined by
     $\delta_{1}+\delta_{2}=0$ is instrumental and attributed to
     higher order mixing of the signal beams with residual control beam
     light leaking through our polarizing beamsplitter, which near this
     line of degeneracy also contributes to a signal near dc
     frequency. \textbf{(b)} Selected part of the photodiode signal after a light storage
     time of 10 $\mu s$ (solid curve) fitted with a sinusoidal curve (dashed line).
     This modulation arises from the beating between two reaccelerated optical modes
     and is interpreted as evidence for their relative coherence.
     The rectangular shape of the pulse form is determined by the gated signal
     processing of the electronic signal, and the visible spiking
     at the beginning is attributed to
     transients of this electronics.}

     \label{fig:results}
\end{figure}

In initial experiments, we have monitored the preparation of dark
states in our experimental configuration using cw signal and
control fields. Fig. \ref{fig:darkres} shows (averaged) optical
transmission spectra of the total signal field as a function of
the signal beam frequencies, both of which were independently
tuned. The signal beam power is 130 $\mu W$ per beam and the
control beam power is 270 $\mu W$. We observe an increased
transmission when either the two-photon detuning
$\delta_{1}=\omega_{C} - \omega_{S1} +\frac{\mu_{B}B}{2}$ or
$\delta_{2}=\omega_{C} - \omega_{S2} -\frac{\mu_{B}B}{2}$
approaches zero, in which case $\sigma_{+}-\pi$ ($\sigma_{-}-\pi$)
polarized optical fields probe whether there is destructive
interference of excitation amplitudes. As expected, the overall
transmission signal is maximized near $\delta_{1} = \delta_{2} =
0$, the region of tripod-type transparency with two dark states.
We wish to point out that tripod-type dark resonance physics has
also been investigated in
\cite{PhysRevA.54.1556,PhysRevLett.91.213001,Meshulam:07}. In
other preparatory experiments, we observed slow light by pulsing
the signal beams. Typical observed group velocities were 1700
$m/s$ with the tripod system.

We then moved to a storage of light in the tripod polariton
system. The input signal pulses typically used were rectangularly
shaped and had a temporal length of 20 $\mu s$. The intensity of
the control beam is ramped to zero at the falling edge of the
signal beams pulse. After a storage time of typically 10 $\mu s$,
only the control beam was pulsed on. Fig. \ref{fig:storedlight}
gives a typical result obtained when storing a pulse performed
with both signal fields simultaneously. The plot shows the
obtained photodiode signal, onto which both signal beams are
imaged behind the cell, after gated signal processing. We observe
a slow decay in the retrieved signal beams pulse, on which a
sinusoidal modulation is visible. We attribute this modulation to
the beating of the two retrieved signal beams fields. From this
result we conclude that simultaneous storage of two signal fields
is possible and that their relative phase information is
preserved. Note that here, the photodiode signal was
electronically shut off by a switch during the storage period and
processed in the part of the read-out period where the amplitude
of the beat signal was sufficiently large. Due to this detection
technique the depicted shape of the read-out pulse is rectangular
rather than exponentially decaying to zero.

Using the tripod scheme, we can exploit the full two-dimensional
subspace that is dark for the light field in this configuration.
Qualitatively, the experiment can be understood by considering the
simplified level scheme of Fig. 1. By shutting off both signal
beams and the control beam adiabatically, the state of the
propagating input light field is mapped into a spin-wave coherence
in the tripod system with two degrees of freedom. When the signal
beam pulses are contained within the medium, the coherence of the
laser fields is already imprinted on the atoms. As the control
laser is turned back on, the signal laser pulses are regenerated
by adiabatic following mapping atomic coherences again into the
corresponding signal field modes. The pulses continue to propagate
under tripod-EIT conditions as prior to the turn-off of the
control beam. The interference observed between the two released
signal fields shows their relative coherence, which we interpret
as the beating of two polariton modes. Note that in contrast to
the free photon case, where the spin direction is parallel to the
propagation direction, in our experimental configuration for each
of the polariton modes the atomic contribution of the spin wave
state is directed transversely to the propagation direction.

\begin{figure}[h!]
  \begin{center}
    \includegraphics [width=0.4\textwidth]{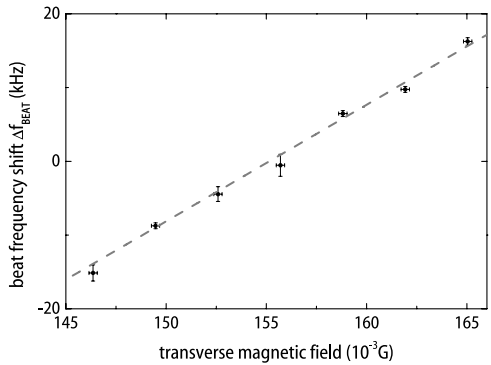}
  \caption{\label{fig:beat}  Measured beat frequency shift of the released signal beams, after storage as a function of the applied (transverse) magnetic field. For all shown data points, the atomic sample was irradiated with optical signal fields of the same frequency during the storage procedure.}
  \end{center}
 \end{figure}

In subsequent experiments, we have varied the magnetic bias field
over a region within the spectral width of the EIT resonance. Fig.
\ref{fig:beat} shows the difference between the measured beat
frequency and the incident signal beams difference frequency as a
function of the magnetic bias field. We observe a linear
dependence of the beat frequency on the magnitude of the magnetic
field. Within our experimental accuracy, the beat frequency equals
the Zeeman splitting $2 g_F\mu_{B}B$ of a $\Delta m_{F}=2$ Raman
coherence in the used rubidium $F=2$ ground state system, where
$g_F$ is the (hyperfine) g-factor that equals 1/2 here. Note that
all shown data points have been recorded with the same signal
beams difference frequency. That is to say, while the storage of
light can be done with frequency difference somewhat differing
from the atomic (Raman) transition frequency, as long as one is
within the EIT transparency window, the retrieved optical
difference frequency within our experimental accuracy is determined by the atomic transition frequency of the
ground states $\left|g_{-}\right\rangle$ and
$\left|g_{+}\right\rangle$ of the tripod coherence (i.e. the
Zeeman states $m_{F}$ and $m_{F}+2$ of the rubidium level scheme).
This result is expected, since the storage onto a spin wave
coherence imprints phase and amplitude information. Further, the regenerated optical dipole for a $\sigma_+$
($\sigma_-$) transition oscillates at a frequency $\omega_{C}\pm
g_F\mu_{B}B$, where $\omega_{C}$ denotes the control beam
frequency and $g_F\mu_{B}B$ the Zeeman splitting between two
adjacent magnetic ground state sublevels. The difference frequency
between the two retrieved signal beams is thus expected to be $2
g_F\mu_{B}B$. The spinor polariton system acts as an atomic
frequency converter, i.e. coherences stored in the system at a
particular signal beam difference frequency after
storage are converted to the atomic (signal beams) transition
frequency.

We conjecture that these experiments pave the way towards a
storage of quantized signal beams states, where e.g. squeezed
signal light states would be of particular interest. Since the
light storage despite its frequency conversion to the atomic
transition frequency is expected to conserve the quantum
properties, we expect that measurements of the retrieved signal
beams difference are possible with sub-shot noise precision,
allowing for quantum limited measurements of magnetic fields and
atomic transition frequencies. We anticipate that with a suitable
mechanism for driving the internal two-level structure of the
spinor polariton an all-optical quantum frequency switch can be
implemented.

To conclude, we have stored light in an ensemble of tripod-type
atoms, demonstrating the mapping of dark polaritons with internal
two-level structure into spin-wave coherences and their coherent
retrieval. Also, we have shown that the retrieved difference
frequency of the emitted signal beams is determined by the atomic
transition frequency of ground state Zeeman sublevels.

For the future, we anticipate applications of dark polaritons with
internal structure in the fields of quantum limited precision
metrology and quantum information.

\begin{acknowledgments}
We acknowledge support by the SFB/TR 21 of the Deutsche Forschungsgemeinschaft.
\end{acknowledgments}

\begin{thebibliography}{19}
\bibitem{loss_comm}
These loss channels could be avoided by using ultracold rubidium
atoms, for which the signal beams could be irradiated in a
direction perpendicular to the control beam axis, and along the
magentic bias field so that pure $\sigma_{+}$ and $\sigma_{-}$
polarizations are possible. For the thermal sample used in the
present work such a non-collinear beam configuration cannot be
used due to the requirement of a Doppler-free scheme.
\end{thebibliography}
\end{document}